\documentclass[aps,amsmath,prd,preprint,floats,epsf,superscriptaddress,nofootinbib]{revtex4}
\pdfoutput = 1
\usepackage{graphicx}
\usepackage{bm}
\usepackage{color}
\usepackage{dcolumn}

\begin{document}

\preprint{IPMU 17-0147}
\bigskip

\title{Current Status of Top-Specific Variant Axion Model}

\author{Cheng-Wei Chiang}
\email[e-mail: ]{chengwei@ncu.edu.tw}
\affiliation{Department of Physics, National Taiwan University, Taipei, Taiwan 10617, R.O.C.}
\affiliation{Institute of Physics, Academia Sinica, Taipei, Taiwan 11529, R.O.C.}
\affiliation{Kavli IPMU (WPI), UTIAS, The University of Tokyo, Kashiwa, Chiba 277-8583, Japan}

\author{Hajime Fukuda}
\email[e-mail: ]{hajime.fukuda@ipmu.jp}
\affiliation{Kavli IPMU (WPI), UTIAS, The University of Tokyo, Kashiwa, Chiba 277-8583, Japan}

\author{Michihisa Takeuchi}
\email[e-mail: ]{michihisa.takeuchi@ipmu.jp}
\affiliation{Kavli IPMU (WPI), UTIAS, The University of Tokyo, Kashiwa, Chiba 277-8583, Japan}

\author{Tsutomu T. Yanagida}
\email[e-mail: ]{tsutomu.tyanagida@ipmu.jp}
\affiliation{Kavli IPMU (WPI), UTIAS, The University of Tokyo, Kashiwa, Chiba 277-8583, Japan}
\affiliation{Hamamatsu Professor}

\date{\today}

\begin{abstract}
The invisible variant axion model is very attractive as it is free from the domain wall problem.  This model requires at least two Higgs doublets at the electroweak scale, with one Higgs doublet carrying a nonzero Peccei-Quinn (PQ) charge and the other being neutral under the PQ $\text U(1)$ symmetry.  We consider a scenario where only the right-handed top quark is charged under the PQ symmetry and couples with the PQ-charged Higgs doublet.  As a general prediction of this model, 
the top quark can decay to the observed standard model-like Higgs boson $h$ and the
charm or up quark, $t\to h~ c/u$, which recently exhibit slight excesses at LHC Run-I and Run-II. 
It will soon be testable at the LHC Run-II. 
If the  rare top decay excess stays at the observed central value, we show that $\tan \beta \sim 1$ or smaller is preferred by the Higgs data. The chiral nature of the Higgs flavor-changing interaction is a distinctive feature, and can be tested using the angular distribution of the $t \to ch$ decays at the LHC.
\end{abstract}


\maketitle

\section{Introduction } 
\label{sec:intro}

The strong CP problem can be elegantly solved by the Peccei-Quinn (PQ) mechanism\,\cite{Peccei:1977hh}, where a $\text{U}(1)_{PQ}$ symmetry is employed to rotate away $\theta_{\rm QCD}$, the CP-violating phase in QCD.  Not manifest in the standard model (SM), the PQ symmetry must be broken spontaneously, thereby predicting the existence of a Nambu-Goldstone boson.  Since the PQ symmetry is anomalous, the additional light degree of freedom associated with the symmetry breaking is a massive pseudo Nambu-Goldstone boson, the axion\,\cite{Weinberg:1977ma,Wilczek:1977pj}.  Dynamics of the axion is characterized by the axion decay constant $f_a$.  The lower bound on $f_a$ is obtained from axion helioscopes and astronomical observations to be $f_a \gtrsim 10^9$\,GeV (see, for example, Ref.\,\cite{Agashe:2014kda}).  Moreover, coherent oscillations of the axion field can play the role of cold dark matter in the present Universe\,\cite{Abbott:1982af,Preskill:1982cy,Dine:1982ah}, from which one determines $f_a \sim 10^{10-12}$\,GeV\,~\cite{Ade:2015xua,Kawasaki:2014sqa}. if the axion is the dominant component of dark matter.
This beautiful mechanism, however, suffers from the problem of domain wall formation in the early Universe.  This is because the model has $N_{DW}=3$ discrete vacua related to the number of fermion families.

The variant axion model introduced in Refs.~\cite{Peccei:1986pn,Krauss:1986wx} is an interesting axion model as it is free from the above-mentioned domain wall problem.  This is achieved by allowing only one right-handed quark to carry a PQ charge and thus rendering a unique vacuum $(N_{DW}=1)$\,\cite{Geng:1990dv}.  For consistency, the model requires two Higgs doublet fields, one of which is also charged under the PQ symmetry.  As a result, there is a non-trivial flavor structure in the Yukawa couplings\,\cite{Chen:2010su} that can lead to flavor-changing neutral-current (FCNC) couplings of the Higgs bosons to at least quarks.  Besides, such FCNC couplings depend on the chirality of fermions not seen in the common two-Higgs doublet models (2HDM's).   Therefore, the variant axion model exhibits interesting and distinctive phenomena in flavor physics at low energies.

In Refs.~\cite{Chen:2010su,Chiang:2015cba}, we had considered such a 2HDM with the PQ symmetry and assign a nonzero charge to the right-handed top quark, thus dubbed the top-specific variant axion model. Previously, we had performed the parameter scan based on the constraints on the mixing parameters in the Higgs sector using the LHC Run-I data~\cite{Chiang:2015cba}.  In particular, there was an interesting excess in the $h\to \tau \mu$ decay\,\cite{Khachatryan:2015kon, Aad:2015gha} at that time.  Therefore, we had focused on the compatibility of this model to the excess and discussed the connection between the quark sector and the lepton sector. Recently, the new Run-II results negate the excess and have a result being well consistent with the SM prediction~\cite{CMS:2017onh}.  In this letter, we aim to update the parameter fitting using the current Higgs data, and study its primary signature of the $t \to h~ c/u$ decays that are being probed at the LHC.

Recently, the ATLAS Collaboration has reported a slight excess in the $t \to ch$ channel at $1.5\sigma$ level, $BR(t \to ch) = (0.22 \pm 0.14)$\%, using the full set of the LHC Run-I data~\cite{Aad:2015pja}.
That excess is consistent with the corresponding CMS observation at LHC Run-I, given by $BR(t \to ch) < 0.40\%$ (or $BR(t \to uh) < 0.55\%$) at 95\% confidence level (C.L.)~\cite{Khachatryan:2016atv}.
More recently, an ATLAS analysis based on the LHC Run-II data is published and also suggests an interesting excess in the $t \to c h$ decay~\cite{Aaboud:2017mfd}, where the observed results set an upper bound $BR(t\to ch)<0.22\%$ at 95\% CL, while the expected one is $BR(t\to ch)<0.16 \%$, which corresponds to a $1 \sigma$ excess.
Although currently the data seem to merely show a small upper fluctuation, it may turn out to be revealing a real new physics signature in the long run.  It is thus worth discussing whether the variant axion model is compatible with such an excess at this point, since such a FCNC top decay is one of the robust and distinctive predictions of the model.

In this paper, we take the central value of $BR(t\to ch)=0.22\%$ given by the ATLAS in Run-I as a nominal value (also the 95\% CL upper bound of the ATLAS Run-II determination).  It corresponds to $\lambda_{tcH}=0.090$ or $a^2 \sin^2\rho = (2.2\times 10^{-3})/(3.24\times 10^{-2})=0.068$ in our model parameters.
To obtain such a sizable effect, large mixing in $\rho$ or large $a=\cos(\beta-\alpha)(\tan\beta + \cot\beta)$ is required. 
Therefore, the off-diagonal couplings have to be sizable and it requires a careful check whether the parameter space is compatible with perturbativity conditions.
By taking the current SM-like Higgs data into account, we set constraints on the parameter space of the model.  
We find that all the constraints are satisfied when $\tan \beta \sim 1$.  Especially, $\tan \beta<1$ is preferred in order to have sizable top FCNC effects without having conflicts with the other Higgs observables and theoretical restrictions.  
We also show how the chirality nature in the FCNC's can be probed by studying the angular distribution of the $t \to c h$ decay at LHC Run-II.

A crude estimate for the future sensitivity on the $t \to c h$ decay without optimization was given 
as $2 \times 10^{-3}(5\times 10^{-4})$ in the lepton channels and $5 \times 10^{-4}(2 \times 10^{-4})$ in the photon channels ~\cite{ATL-PHYS-PUB-2013-012, Agashe:2013hma, ATLAS:2013hta}, 
assuming an integrated luminosity of 300~fb$^{-1}$ (3~ab$^{-1}$) in Run-III. 
In fact, one notes that the current sensitivity already approaches a similar level by combining the $\gamma\gamma$, multi-lepton, and $bb$ modes, with just the LHC Run-I (7 and 8-TeV) data of ${\cal O}(25)$fb$^{-1}$.  We can expect the final combined sensitivity at 3~ab$^{-1}$ to reach below 0.01\%.

This paper is organized as follows. 
Section\,\ref{sec:Lagrangian} discusses the structure of the Higgs sector along with the FCNC couplings of the SM-like Higgs boson to fermions in the top-specific variant axion model. 
We also discuss the expected sizes of FCNC decays and theoretical constraints on this model.
In Section\,\ref{sec:current_bounds}, we perform a $\chi^2$-fit analysis based on the latest Higgs signal strength data at the LHC. 
In Section\,\ref{sec:confirmation}, we discuss how to confirm this model through the angular distribution of $t \to c~ h$ decay and show the expected sensitivity corresponding to the scenario considered in this work. 
Conclusions are given in Section\,\ref{sec:conclusion}.

\section{Top-Specific Variant Axion Model}
\label{sec:Lagrangian}

As a minimal setup of the variant axion model, we introduce two Higgs doublet fields $\Phi_1$ and $\Phi_2$ and a scalar field $\sigma$ with PQ charges $0$, $-1$ and $1$, respectively.  The gauge singlet scalar $\sigma$ gets a vacuum expectation value (VEV) $f_a$ and breaks the PQ symmetry spontaneously at a high energy scale.  It therefore does not play much a role at low energies.  In the quark sector, we assume that only the right-handed top quark field $t_R$ possesses a nonzero PQ charge of $-1$.  Note that we can additionally assign nonzero PQ charges to leptons as well, as they do not contribute to the number of domain walls $N_{DW}$\,\cite{Geng:1990dv}.  In this paper, we focus on the scenario where the leptons have no PQ charges. The interesting phenomenology of the case when the leptons also carry PQ charges can be found in our previous analysis~\cite{Chiang:2015cba}.

The most general renormalizable Higgs potential obeying the PQ symmetry with the above PQ charge assignments is, as already given in Refs.\,\cite{Chen:2010su,Chiang:2015cba,Davidson:2005cw}:
\begin{align}
V(\Phi_1,\Phi_2)
=&
m_{11}^2 \Phi_1^\dagger\Phi_1 + m_{22}^2 \Phi_2^\dagger\Phi_2
- (m_{12}^2 \Phi_1^\dagger\Phi_2 + \mbox{H.c.})
+ \frac{\lambda_1}{2} \left( \Phi_1^\dagger\Phi_1 \right)^2
+ \frac{\lambda_2}{2} \left( \Phi_2^\dagger\Phi_2 \right)^2\nonumber \\
& 
+ \lambda_3 \left( \Phi_1^\dagger\Phi_1 \right)\left( \Phi_2^\dagger\Phi_2 \right)
+ \lambda_4 \left( \Phi_1^\dagger\Phi_2 \right)\left( \Phi_2^\dagger\Phi_1 \right) ~,
\end{align}
where the $\sigma$ field has been integrated out. 
The $m_{12}^2$ terms softly violate the PQ symmetry (as can be derived from a UV-complete Lagrangian~\cite{Chen:2010su}), and can be made real and positive through a rotation of the PQ symmetry.
All the other terms respect the PQ symmetry and their associated parameters ($m_{11}^2$, $m_{22}^2$, and $\lambda_{1,2,3,4}$) are real.

After the electroweak symmetry breaking, each $\Phi_i$ acquires a respective VEV $v_i$ and can be written in terms of component fields as $\Phi_i = (H^+_i, (v_i + h_i + iA_i)/\sqrt{2})^T$.  We define $\tan \beta = v_2/v_1$ as usual and $v^2_{\rm SM} = v_1^2 + v_2^2 \simeq (246~\mbox{GeV})^2$.  
We now rotate the Higgs doublets to the so-called Higgs basis~\cite{Lavoura:1994fv}, where only one of the doublets ($\Phi^{\rm SM}$) 
has a nonzero VEV:
\begin{align}
\begin{split}
&
\begin{pmatrix}
\Phi_1 \\ 
\Phi_2
\end{pmatrix} 
= R_\beta
\begin{pmatrix}
\Phi^{\rm SM}\\ 
\Phi^\prime
\end{pmatrix}
~,~\mbox{with }
R_{\theta} =
\begin{pmatrix}
\cos\theta & -\sin\theta \\
\sin\theta & \cos\theta \\
\end{pmatrix} ~,
\\
&
\mbox{and }~
\Phi^{\rm SM} = 
\begin{pmatrix}
G^+ \\
(v_{\rm SM} + h^{\rm SM} + iG^0)/\sqrt{2}\\
\end{pmatrix}~,~ 
\Phi^{\prime} = 
\begin{pmatrix}
H^+ \\
(h^{\prime} + iA^0)/\sqrt{2}\\
\end{pmatrix} ~,
\end{split}
\end{align}
where $G^\pm$ and $G^0$ are the would-be Nambu-Goldstone bosons to become the longitudinal modes of the $W^\pm$ and $Z$ bosons.  The pseudoscalar Higgs boson $A^0$ and charged Higgs boson $H^\pm$ are mass eigenstates with masses $m_A$ and $m_{H^+}$, respectively.  We also define the mass eigenstates of the CP-even neutral Higgs bosons as $h$ and $H$, with respective masses $m_h$ and $m_H$ ($m_h < m_H$), through a rotation of angle $\alpha$ from the original PQ basis as follows:
\begin{align}
\begin{pmatrix}
H\\
h 
\end{pmatrix}
=
R_{-\alpha} 
\begin{pmatrix}
h_1 \\
h_2 \\
\end{pmatrix}
=
R_{ \beta - \alpha } 
\begin{pmatrix}
h^{\rm SM} \\
h^\prime \\
\end{pmatrix} ~.
\end{align}
Note that the light Higgs boson $h$ becomes a SM-like Higgs boson $h^{\rm SM}$ in the limit of $\sin(\beta - \alpha) \to 1$, which can be realized when $m_{12}^2 \to \infty$. 
 The couplings between $h$ and weak gauge bosons are read as
\begin{align}
g_{hVV}=\sin(\beta-\alpha) g_{hVV}^{\rm SM} ~,\ \ \ 
g_{HVV}=\cos(\beta-\alpha) g_{hVV}^{\rm SM} ~,\ \ \ {\rm and}\ \ \  g_{AVV}=0 ~,
\label{eq:HVV}
\end{align}
where $g_{hVV}^{\rm SM}$ are the couplings in the SM.  
The triple Higgs coupling $\lambda_{hH^+H^-}$, defined by the $\lambda_{hH^+H^-}{hH^+H^-}$ interaction term in the Lagrangian, is given by:
\begin{align}
\begin{split}
\lambda_{hH^+H^-}
&= \frac{1}{v_{\rm SM}}
\left[
(m_h^2 + 2 m_{H^+}^2 - 2 m_{A}^2) \sin(\beta- \alpha) 
\right.
\\
& \qquad\qquad \left. 
+ (m_A^2 + m_h^2) (\tan\beta - \cot \beta )\cos (\beta - \alpha)
\right] ,
\end{split}
\label{eq:HHH}
\end{align}
where $m_A^2 \equiv 2m_{12}^2 / \sin2\beta$.

For Yukawa interactions, since we consider the scenario in which only the right-handed top quark carries a nonzero PQ charge among all quark fields, the up-type Yukawa interaction Lagrangian is:
\begin{eqnarray}
\mathcal L^u = -\Phi_1 \overline{u}_{R a} [Y_{u1}]_{ai} q_{i} - \Phi_2 \overline{u}_{R 3} [Y_{u2}]_{i} q_{i} + \text{H.c.}
\end{eqnarray} 
where the family indices $a = 1,2$ and $i, j= 1,2,3$. 
Schematically, the Yukawa coupling matrices, $Y_{u1}$ and $Y_{u2}$, in the original PQ basis take the forms:
\begin{eqnarray}
Y_{u1}=\begin{pmatrix}
* &* &*\\
* &* &*\\
0 &0 &0 \\
\end{pmatrix}~,~~ 
Y_{u2}=\begin{pmatrix}
0 &0 &0 \\
0 &0 &0 \\
* &* &*\\
\end{pmatrix}~,
\end{eqnarray} 
where $*$ indicates a generally nonzero element.

In the Higgs basis, the up-type Yukawa interaction Lagrangian can be expressed as
\begin{eqnarray}
\mathcal L^u = -\Phi^{\rm SM} \overline{u}_{R i} [Y_u^{SM}]_{ij} q_{j} - \Phi^\prime \overline{u}_{R i} [Y_u^\prime]_{ij} q_{j} + \text{h.c.} ~,
\end{eqnarray} 
and the Yukawa matrices are 
\begin{align}
\begin{split}
Y_u^{\rm SM}  &= \cos{\beta} Y_{u1} + \sin{\beta} Y_{u2}~,
\\
Y_u^\prime &= -\sin{\beta} Y_{u1} + \cos{\beta} Y_{u2} 
= \begin{pmatrix}
-\tan\beta &&\\
& -\tan\beta &\\
&& \cot\beta \\
\end{pmatrix} Y_u^{SM} ~.
\end{split}
\end{align} 
Therefore, the number of degrees of freedom in the up-type Yukawa matrices of this model is the same as that of the SM.
At this stage, the mass matrix $M_u\equiv\frac{v_\text{SM}}{\sqrt{2}}Y_u^{\rm SM}$
is generally non-diagonal, and can be brought to its diagonal form through a bi-unitary transformation $V M_u U^\dagger=\text{diag}(m_u,m_c,m_t)\equiv \frac{v_\text{SM}}{\sqrt2}Y_u^\text{diag}$, where $U$ and $V$ are two unitary matrices,
which rotates the left-handed fields $q_i$ and the right-handed fields $u_{R,i}$, respectively.   
In this basis, the other Yukawa matrix becomes
\begin{eqnarray}
\begin{split}
Y_u^{\prime, \text{diag}}
&=  \begin{pmatrix}
-\tan\beta &&\\
& -\tan\beta &\\
&& \cot\beta \\
\end{pmatrix} Y_u^\text{diag} + (\tan\beta + \cot\beta)H_u
 Y_u^\text{diag},
\end{split}
\label{eq:Yuprime}
\end{eqnarray}
where the Hermitian matrix
\begin{align}
H_u\equiv V \begin{pmatrix}
0 &&\\
& 0 &\\
&&  1  \\
\end{pmatrix} V^\dagger - \begin{pmatrix}
0 &&\\
& 0 &\\
&&  1  \\
\end{pmatrix} ~.
\end{align}
Note that in the second term of Eq.~(\ref{eq:Yuprime}), the $(\tan\beta + \cot\beta)H_u$ part describes  mixing among up-type quarks and $Y_u^\text{diag}$ controls the strength of coupling with the dominant component given by the top Yukawa coupling.  For simplicity, we will omit the superscript ``diag'' while working in the mass-diagonal basis in the following discussions.  Note that $V$ is the rotation matrix for the right-handed up-type quarks and is 
completely independent of the CKM matrix, which is the product of left-handed up-type and left-handed down-type quark rotation matrices.  Therefore, the mixing angles in $V$ can be as large as $\mathcal O(1)$, a key intriguing feature of the model.

As an illustration and in anticipation of interesting collider phenomenology associated with the top quark, we restrict ourselves to the case of $t$-$c$ mixing in this paper as in Ref.~\cite{Chiang:2015cba}. 
In such a simplified scenario without introducing new CP phases, 
the mixing matrices $H_u$ and $V$ can be parameterized in terms of $\rho$ only as:
\begin{eqnarray}
\label{eq:mixing_parametrization}
H_u=
\begin{pmatrix}
0 & 0 & 0 \cr
0 & \frac{1 - \cos\rho}{2}  &   \frac{\sin\rho}{2}  \\
0 &\frac{\sin\rho}{2}  & \frac{\cos\rho -1}{2}  \\
\end{pmatrix} ~,\ 
V = \begin{pmatrix}
1 & 0 & 0 \\
0 & \cos\frac\rho2 & \sin\frac\rho2 \\
0 & -\sin\frac\rho2 & \cos\frac\rho2 \\
\end{pmatrix} ~,\ 
\end{eqnarray}
and the Yukawa interactions of the observed Higgs boson $h$ in the mass eigenbasis are then described by
\begin{align}
\label{eq:Yukawa}
\mathcal L_Y &\equiv 
- \sum_{f=e,\cdots,u,\cdots,d,\cdots} \xi_f \frac{m_f}{v_\text{SM}} h \overline{f}f + \mathcal{L}_\text{FCNC}
\\
& \mbox{with }~
\label{eq:Yukawa_FCNC}
\mathcal L_\text{FCNC} = 
\mathcal L_{tc} =
-\frac{a\sin\rho}{2v_\text{SM}}
(m_t \bar{c}_R  t_L + m_c \bar{t}_R c_L)h
+ \text{H.c.}\,
\end{align}
where $a \equiv (\tan\beta +\cot\beta) \cos(\beta - \alpha)$ and,
\begin{eqnarray}
\xi_f&=&\begin{cases}
\,\sin(\beta - \alpha) + \left(\cot\beta - \frac{1-\cos\rho}{2}(\tan\beta + \cot\beta)\right)\cos(\beta-\alpha) & \mbox{for } f=t
~, \\
\,\sin(\beta-\alpha) - \left(\tan\beta - \frac{1-\cos\rho}{2}(\tan\beta + \cot\beta)\right)\cos(\beta-\alpha) & 
\mbox{for } f=c
~,\\
\,\sin(\beta - \alpha) - \tan\beta \cos(\beta - \alpha) & \mbox{otherwise}~.
\end{cases} 
\label{eq:xi2}
\end{eqnarray}


One striking feature of $\mathcal L_\text{FCNC}$ is that the predicted flavor violation is associated 
with large asymmetries in the chirality.  In this simplified case, the top decay is dominated by the right-handed charm-associated processes to be discussed in more detail in the next section.

\bigskip

\subsection{Rare Top quark FCNC decay}

This model generically predicts the top FCNC decay $t \to ch$ (or $t \to uh$) via the mixing effect.  Therefore, such decays serve as a smoking gun signature of the model.  
For definiteness, we focus on the $t \to ch$ decay in this section, but note that the current experimental limits do not actively tag the flavor of the accompanied jet and that what is constrained is the weighted sum of all branching ratios of $t \to qh$ ($q = u, c$).

The partial decay width of $t\to ch$ is given by
\begin{align}
\begin{split}
\Gamma_{t \to ch}
&= \frac{G_F m_t^3  a^2  \sin^2\rho}{64 \pi \sqrt{2}} \left( 1 - r_h^2 \right)^2,
\end{split}
\end{align}
with $r_h^2 \equiv m_h^2/m_t^2 \sim 0.522$ for $m_h=125$~GeV and $m_t=173$~GeV.  
By comparing it with the width of $t \to bW$ in the SM at the leading order,
\[
\Gamma_{t\to bW} = \frac{G_F |V_{tb}|^2 m_t^3}{8\pi \sqrt{2}}
\left( 1 - r^2_W \right)^2 \left( 1 + 2r^2_W \right) ~,
\]
with $r_W^2\equiv m_W^2 / m_t^2 \simeq 0.214$ for $m_W=80.4$~GeV,
we can obtain by assuming $BR(t \to bW)$ is close to unity that
\begin{align}
BR( t \to c h) \simeq \frac{\Gamma_{t\to ch}}{\Gamma_{t\to bW}} \simeq 
(3.24 \times  10^{-2} )  a^2  \sin^2\rho ~.
\end{align}
The nominal branching ratio of 0.22~\% corresponds to the mixing parameters satisfying
\[ 
a^2 \sin^2\rho = 0.068 ~.
\]
As the future sensitivity of $0.02$\% for $BR( t \to c h)$ (for 14~TeV and 3000~fb$^{-1}$) corresponds to 
\[ 
a^2 \sin^2 \rho = 6.2\times 10^{-3} ~,
\]
the current nominal value will be fully confirmed at $5\sigma$ level by then if the model is correct.
These two values will be depicted in the following figures in pink and red, respectively.

It is noted that the $h$-$t$-$c/u$ couplings can also contribute to other flavor observables. 
For example, $D^0$-$\bar{D^0}$ meson mixing measurements constrain the products
$|\lambda_{tu} \lambda_{tc}|,|\lambda_{ut} \lambda_{ct}| < 7.6\times10^{-3}$,
$|\lambda_{tu} \lambda_{ct}|,|\lambda_{ut} \lambda_{tc}| < 2.2\times10^{-3}$ and 
$|\lambda_{tu} \lambda_{ut}\lambda_{ct} \lambda_{tc}|^{1/2} < 0.9\times10^{-3}$~\cite{Harnik:2012pb,Bona:2007vi}. 
As we are mainly concerned with the mixing between the last two generations, 
combining with $|\lambda_{ct}|=|a\frac{m_t}{v_{\rm SM}} (H_u)_{ct}|= 0.09$ coming from the nominal value of $BR(t\to hc)$, 
the most stringent constraint is $|\lambda_{ut} \lambda_{ct}| = |a^2 \frac{m_t^2}{v_{\rm SM}^2} (H_u)_{ut}(H_u)_{ct}| \le 0.0086 |(H_u)_{ut}/(H_u)_{ct}|$.
Thus, the condition $|(H_u)_{ut}/(H_u)_{ct}| \lesssim 0.9$
is sufficient to avoid the above-mentioned constraints while reproducing the nominal value 
with no significant fine-tuning in the model parameters.
While introducing no new CP-violating sources in this model, we note in passing that the imaginary parts of the flavor-violating Yukawa couplings can be constrained by the hadron electric dipole moments and CP-violating observables in the $D$ mesons~\cite{Gorbahn:2014sha}.

\subsection{Perturbativity constraints}

Since $Y_u^{\prime, \text{diag}}$ involves $\rho$ and $\tan\beta$, some element may become too large for some sizeable $\rho$, $\tan\beta$ or $\cot\beta$.  Large Yukawa couplings could have the problem of blowing up after running to high energies.
Since our model assumes the PQ symmetry to solve the strong CP problem, the coupling must not blow up at least up to the PQ scale.  More stringently, we require that the theory does not contain any divergent coupling up to the Planck scale.  As the Yukawa couplings are base-dependent, we require that the absolute value of any Yukawa coupling is smaller than $4\pi$ for the validity of perturbation.

\bigskip
\begin{figure}[h]
\includegraphics[width=0.45\linewidth]{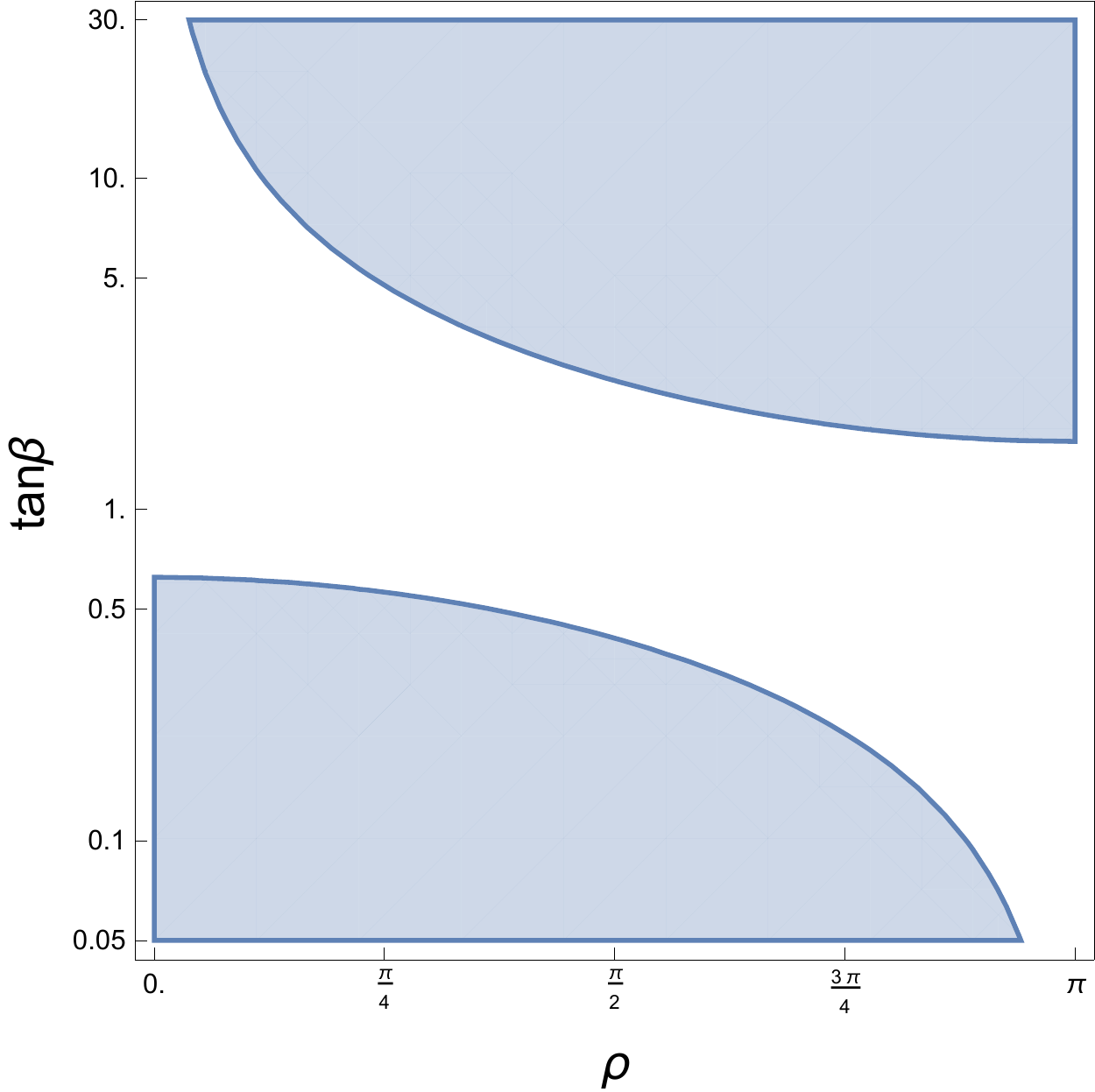}
\caption{The perturbative parameter space in $\rho-\tan\beta$ plane. Shaded regions are excluded by the perturbativity condition ($\forall_{i,j}$ $Y_{ij}(\mu) <4\pi$ at any $\mu<M_{\rm Planck}$ ).}
\label{fig:perturbative}
\end{figure}

We use the 1-loop renormalization equations in Eqs.~(404) to (409) of Ref.~\cite{Branco:2011iw}, ignoring the Higgs self couplings, for numerical evaluations.
The parameter region satisfying the above perturbative condition is shown in Fig.~\ref{fig:perturbative}, with the excluded region shown in gray. 
The result can be understood as follows.
For $\rho = 0$, $\left( Y_u^{\prime, \text{diag}} \right)_{tt} \simeq y_t \cot\beta$ blows up when $\tan\beta$ is small. 
On the other hand, for $\rho\sim \pi$,
$\left( Y_u^{\prime, \text{diag}} \right)_{tt} \simeq y_t \tan\beta$ becomes large when $\tan\beta$ is large.
For the large mixing angle region, $\rho\sim\pi/2$, both large and small $\tan\beta$ cause the coupling to blow up, due to the large off-diagonal element:
\begin{eqnarray}
\left( Y_u^{\prime, \text{diag}} \right)_{ct} = (\tan\beta + \cot \beta)\frac{\sin\rho}{\sqrt{2}}\frac{m_t}{v}.
\end{eqnarray}
Therefore, a moderate $\tan\beta \sim 1$ is preferred by the perturbativity
for a large mixing of $\rho \sim \pi/2$. 
Constraints for several fixed values of $\rho$ are given as follows:
$\tan\beta>0.62$ ($\rho=0$), 
$0.41 < \tan\beta<2.45$ ($\rho=\pi/2$), 
$0.21 < \tan\beta<1.78$ ($\rho=3\pi/4$), 
and $\tan\beta<1.60$ ($\rho=\pi$).
Note that when $\rho$ takes a finite value, the lower bound on $\tan\beta$ is relaxed compared with the usual $\rho=0$ case.
When we fix $\tan\beta$ to a few values, the constraints are: $1.15<\rho$ ($\tan\beta=0.5$)
$\rho<1.99$ ($\tan\beta=2$),
$\rho<0.75$ ($\tan\beta=5$),
and $\rho<0.37$ ($\tan\beta=10$).
We find that there are no constraints on $\rho$ for $0.62<\tan\beta<1.6$.

\section{Higgs signal strength constraints from LHC Run-II}
\label{sec:current_bounds}

\begin{table}[t]
\scriptsize
\begin{ruledtabular}
\begin{tabular}{cccccc}
 & $\gamma\gamma$ & $ZZ$ & $WW$ & $\tau\tau$ & $bb$
\\
\hline
ggF$^{\rm 7,8TeV}$
& $1.10^{+0.23}_{-0.22}$ & $1.13^{+0.34}_{-0.31}$ & $0.84 \pm 0.17$ & $1.0 \pm 0.6$ & --
\\
ggF$^{\rm 13TeV}_{\rm ATLAS}$ & $0.8^{+0.19}_{-0.18}$~\cite{ATLAS:2017myr} & $1.11\pm 0.245 $~\cite{ATLAS:2017cju} & -- & -- & --
\\
ggF$^{\rm 13TeV}_{\rm CMS}$ & $1.11^{+0.19}_{-0.18}$~\cite{CMS:2017rli} & $1.20^{+0.22}_{-0.21}$~\cite{CMS:2017jkd} &
$0.9^{+0.4}_{-0.3}$~\cite{CMS:2017pzi} & 
$1.17^{+0.47}_{-0.40}$~\cite{Sirunyan:2017khh}
&  $2.3 ^{+1.8}_{-1.6}$~\cite{CMS:2017cbv}
\\
\hline
VBF$^{\rm 7,8TeV}$
& $1.3 \pm 0.5$ & $0.1^{+1.1}_{-0.6}$ & $1.2 \pm 0.4$ & $1.3 \pm 0.4$ & --
\\
VBF$^{\rm 13TeV}_{\rm ATLAS}$ & $2.1 \pm 0.6$~\cite{ATLAS:2017myr} & $4.0\pm 1.77$~\cite{ATLAS:2017cju} & $3.2^{+4.4}_{-4.2}$~\cite{ATLAS:2016gld}& -- & $-3.9^{+2.8}_{-2.7}$~\cite{ATLAS:2016lgh}
\\
VBF$^{\rm 13TeV}_{\rm CMS}$ & $0.54^{+0.6}_{-0.5}$~\cite{CMS:2017rli} & $0.06^{+1.03}_{-0.06}$~\cite{CMS:2017jkd} & 
$1.4\pm 0.8$~\cite{CMS:2017pzi}
 & $1.11^{+0.34}_{-0.35}$~\cite{Sirunyan:2017khh}
&$-3.7^{+2.4}_{-2.5}$~\cite{CMS:2016mmc}
\\
\hline
VH$^{\rm 7,8TeV}$
& $0.5 \pm 1.1$ & -- & $2.3 \pm 1.0 $ & $-0.2\pm 1.1$ & $0.63 \pm 0.3$ 
\\
VH$^{\rm 13TeV}_{\rm ATLAS}$ & $0.7^{+0.9}_{-0.8}$ ~\cite{ATLAS:2017myr} & 
 $<3.8$ ~\cite{ATLAS:2017cju} &$1.7^{+1.1}_{-0.9}$~\cite{ATLAS:2016gld}& -- & $1.20^{+0.42}_{-0.36}$~\cite{ATLAS:2017bic}
\\
VH$^{\rm 13TeV}_{\rm CMS}$ & $2.29^{+1.1}_{-1.0}$ ~\cite{CMS:2017rli} & $<2.8$~\cite{CMS:2017jkd}& 
$-0.3 \pm 1.3$ ~\cite{CMS:2017pzi} & -- &--
\\
\hline
$ttH^{\rm 7,8TeV}$
& $2.2^{+1.6}_{-1.3}$ & -- & $5.0^{+1.8}_{-1.7}$ & $-1.9^{+3.7}_{-3.3}$ & $1.1 \pm 1.0$
\\
$ttH^{\rm 13TeV}_{\rm ATLAS}$ & $0.5 \pm 0.6$~\cite{ATLAS:2017myr} & ${}_{(WW)}$
& $2.5^{+1.3}_{-1.1}$~\cite{ATLAS:2016axz}& ${}_{(WW)}$ & $2.1^{+1.0}_{-0.9}$~\cite{ATLAS:2016axz}
\\
$ttH^{\rm 13TeV}_{\rm CMS}$ & $2.22^{+0.9}_{-0.8}$~\cite{CMS:2017rli} & 
${}_{(WW)}$ & $1.5 \pm 0.5$~\cite{CMS:2017vru} & 
$0.72^{+0.62}_{-0.53}$~\cite{CMS:2017lgc} &
$-0.19 \pm 0.80$~\cite{CMS:2016zbb}
\\
\end{tabular}
\end{ruledtabular}
\caption{Higgs signal strengths of various modes measured at the LHC in 7-, 8-, and 13-TeV collisions.
In the first column, ggF, VBF, VH, $ttH$ refers to the Higgs production mechanisms of gluon-gluon fusion, vpector boson fusion, associated production with gauge bosons ($W/Z$), and 
associated production with $t\bar{t}$, respectively.
The superscript indicates the colliding energy, and the subscript indicates the collaboration.
The first row indicates the final states of the Higgs boson. 
Unless quoted explicitly, the data are generally transcribed from Table~8 of Ref.~\cite{Khachatryan:2016vau}.
In the $ttH$ rows, the numbers shown in $WW$ column is based on  
the analysis in the multi-lepton mode, where all $ZZ/WW/\tau\tau$ modes contribute
although dominated by $WW$ mode. Thus,  $(WW)$ means that 
$ZZ/\tau\tau$ modes are in principle constrained by them, 
while not considered in our analysis.}
\label{tab:sigs}
\end{table}

As we assume that the exotic Higgs bosons are sufficiently heavy to decouple from the low-energy phenomenology~\cite{Chiang:2015cba}, our model is essentially parametrized by only three parameters $\alpha$, $\beta$ and $\rho$.
In this section, we show the constraints on these model parameters using the latest LHC Higgs data.  As noted earlier, the couplings between the SM-like Higgs boson $h$ and the SM particles are modified from their SM values: Eq.~(\ref{eq:HVV}) for the gauge bosons and Eq.~(\ref{eq:xi2}) for the fermions.  We use them to estimate the signal strengths of various Higgs production channels.
We note that the diphoton decay width depends to some extent on the coupling $\lambda_{hH^+H^-}$, which in turn would modify the predicted Higgs signal strengths.  However, such a dependence diminishes under our assumption of heavy exotic Higgs bosons.  Therefore, for definiteness, we set $\lambda_{hH^+H^-}=0$ in the following analysis. 

\begin{figure}[h]
\includegraphics[width=0.45\linewidth]{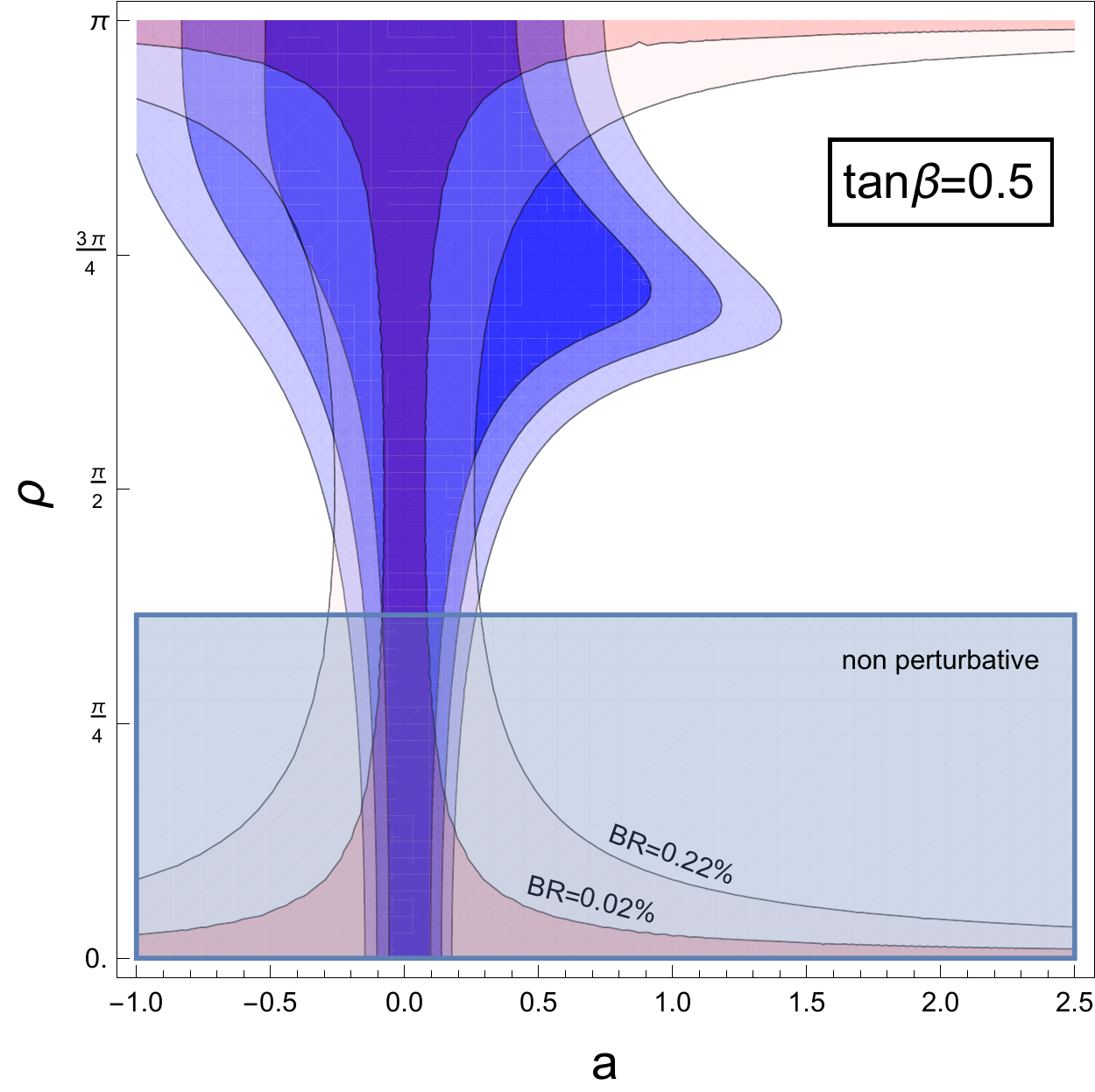}
\hspace{0.5cm}
\includegraphics[width=0.45\linewidth]{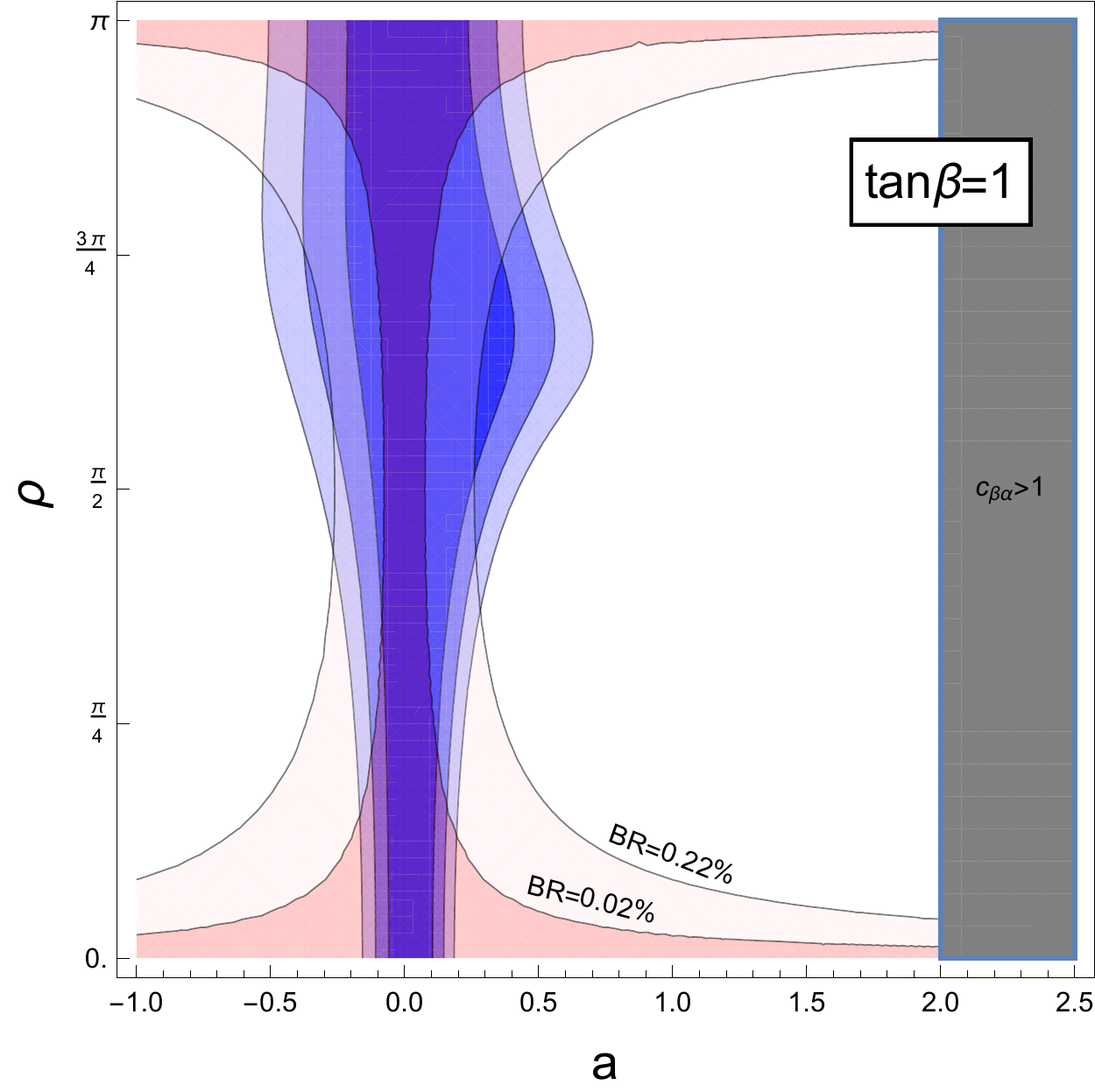}
\\
\bigskip
\includegraphics[width=0.45\linewidth]{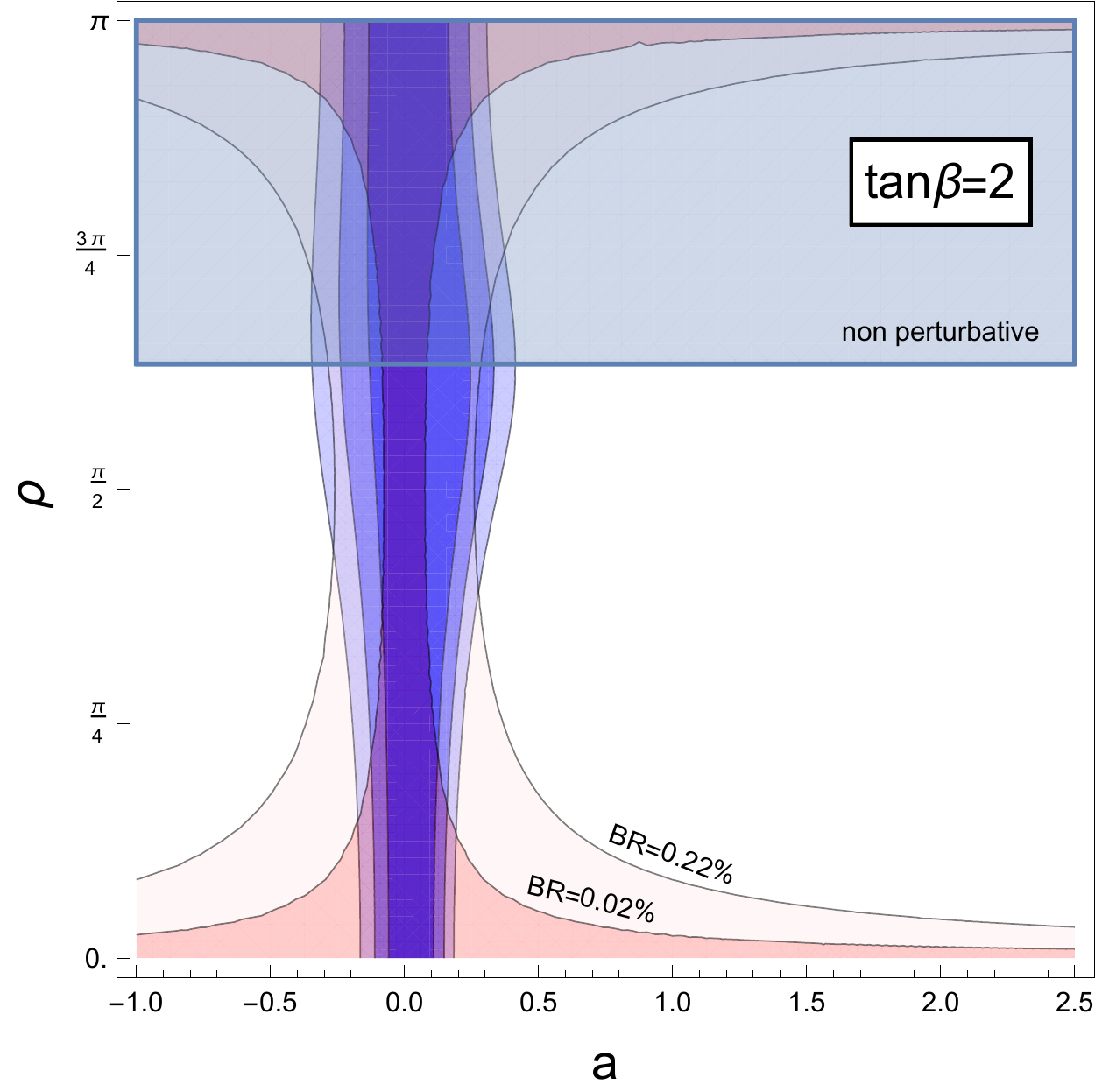}
\hspace{0.5cm}
\includegraphics[width=0.45\linewidth]{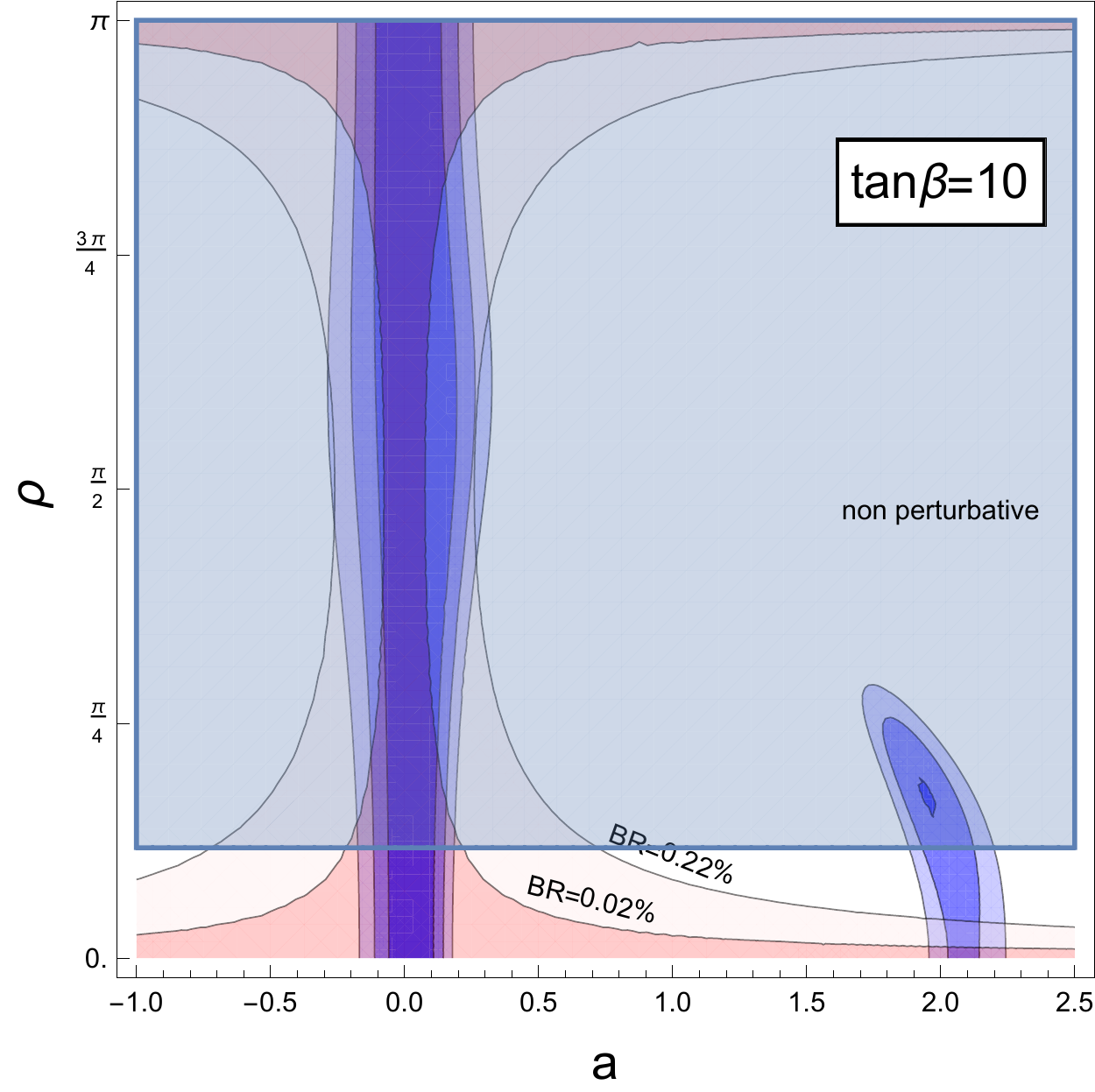}
\caption{Allowed parameter space at $68\%$ (dark blue), $95\%$ (blue) and $99\%$ (light blue) CL from the latest Higgs signal strength data in the $a$-$\rho$ plane for different values of $\tan\beta$: 0.5 (upper-left plot), 1 (upper-right plot), 2 (lower-left plot), and 10 (lower-right plot).
The red contours are drawn for $BR(t \to ch) = 0.22\%$ 
and for the conservative ultimate sensitivity of $0.02\%$.
Assuming the nominal value 0.22\% for $BR(t \to ch)$, $\tan\beta$ of ${\cal O}(1)$ offers a broader parameter space consistent with the current Higgs data at the 1$\sigma$ level, although all $\tan\beta$ value can be made compatible within $2\sigma$.
}
\label{fig:arho}
\end{figure}

In our global $\chi^2$ fit, we take all the signal strengths listed in Table~\ref{tab:sigs} with the both-side errors (44 data) into account. 
They include the Run-I (7-TeV and 8-TeV) data reported by the ATLAS and CMS Collaborations summarized 
in Refs.~\cite{Khachatryan:2016vau} and the latest results from Run 2 at 13 TeV collected from Refs.~\cite{ATLAS:2017myr,ATLAS:2017cju,CMS:2017rli,CMS:2017jkd,
CMS:2017pzi,Sirunyan:2017khh,CMS:2017cbv,ATLAS:2016gld,ATLAS:2016lgh,
CMS:2016mmc,ATLAS:2017bic,ATLAS:2016axz,CMS:2017vru,CMS:2017lgc,
CMS:2016zbb}.
In the $\chi^2$ function, we add the statistical and the systematic errors in quadrature and average asymmetric errors for simplicity.

In Fig.~\ref{fig:arho}, we show the allowed parameter space in the ($a$, $\rho$) plane for $\tan\beta$ fixed to 0.5, 1, 2 and 10. 
The darker blue, blue, and lighter blue regions correspond respectively to the $1\sigma, 2\sigma$ and $ 3\sigma$ regions, based on $\Delta \chi^2 = \chi^2 - \chi^2_{\min} = (3.53,8.02,14.2)$ for 3 degrees of freedom.
We found $\chi^2_{\min} = 48.3$ being an appropriate goodness of fit for 44 observables.
For a fixed $\tan\beta$ value, the blue regions always become broader along the $a$ direction when $\rho$ is turned on and 
broadest around $\pi/2$ to $3\pi/4$. This tendency is stronger when $\tan \beta$ is smaller.
This can be understood as follows from the fact that the latest Higgs data are essentially consistent with the SM expectations, and a large deviation in the signal strengths disfavored.
The gauge boson couplings to the Higgs are consistent with the SM, 
and it forces $\sin(\beta-\alpha) \simeq 1$ and $\cos(\beta-\alpha)$ to be small.
Thus, each signal strength $\mu_i$ from the $ggH$-initiated mode is roughly proportional to $(\xi_t/\xi_b)^2$, 
taking the fact that the total width is mainly controlled by the bottom Yukawa coupling.
From the expression Eq.(\ref{eq:xi2}), we see $d(\xi_t/\xi_b) \simeq (1+\cos\rho)(\tan\beta+ \cot\beta) d\cos(\beta - \alpha)$, 
and this part of the corrections becomes milder with non-zero $\rho$.  Moreover, for small $\tan\beta$, a larger $a$ is realized due to large $\cot\beta$ whereas it does not initiate large effects on $\xi_b$, or on the total width.

There is also so-called ``wrong-sign Yukawa" solution with $\xi_t=1$ and $\xi_i=-1$ ($i$: other than the top) for large $\tan\beta$.  In this case, the Yukawa couplings of quarks other than the top quark have an opposite sign to their SM values~\cite{Ferreira:2014naa}, achieved by having $\tan\beta\cos(\beta-\alpha) \simeq - 2$ but
$(1 - \cos\rho)\tan\beta\cos(\beta-\alpha) \sim 0$.
One can understand why the corresponding solution does not exist for small $\tan\beta$ nor for large $\rho$
from this expression.
For $\tan\beta=10$, there remains viable parameter space at $a\sim 2.2$ as well as normal $a\sim 0$ region 
for small $\rho$ although the wrong-sign region is compatible with the Higgs data at the $2\sigma$ level,
as seen in the right lower plot of Fig.~\ref{fig:arho}.

The FCNC top rare decay process $t \to ch$ can put useful constraints on the parameter space as well.  
Such FCNC effects are proportional to $a^2 \sin^2 \rho$.
The nominal FCNC branching ratios of $\le 0.22~\%$ and 
$\le 0.02$\% are depicted in the plots by the pink and red regions, respectively.
If the nominal size of the signature top decay turns out to be a real signal, 
we can exclude $\rho \sim 0$ region allowed by the current Higgs signal strength data. 

In Fig.~\ref{fig:arho}, we also superimpose the requirement of perturbativity in the Yukawa couplings.  For $\tan\beta > 1$ ($\tan\beta < 1$), larger (smaller) $\rho$ regions are ruled out, as indicated by the shaded light-gray region.  There is no such a constraint for the $\tan\beta = 1$ case.  However, we have a dark-gray region ruled out by $c_{\beta\alpha} > 1$ in this case.

\begin{figure}[h]
\includegraphics[width=0.45\linewidth]{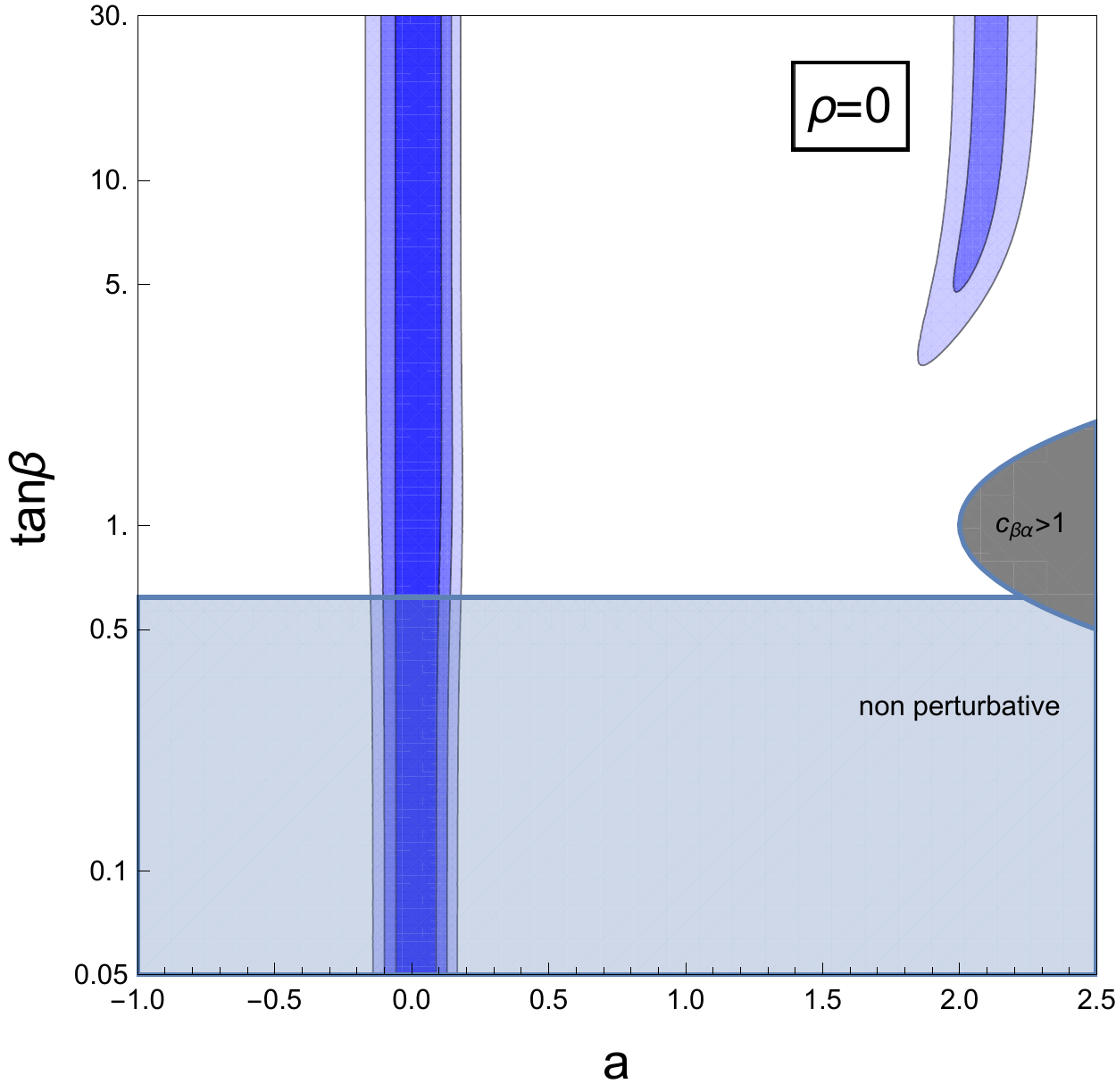}
\includegraphics[width=0.45\linewidth]{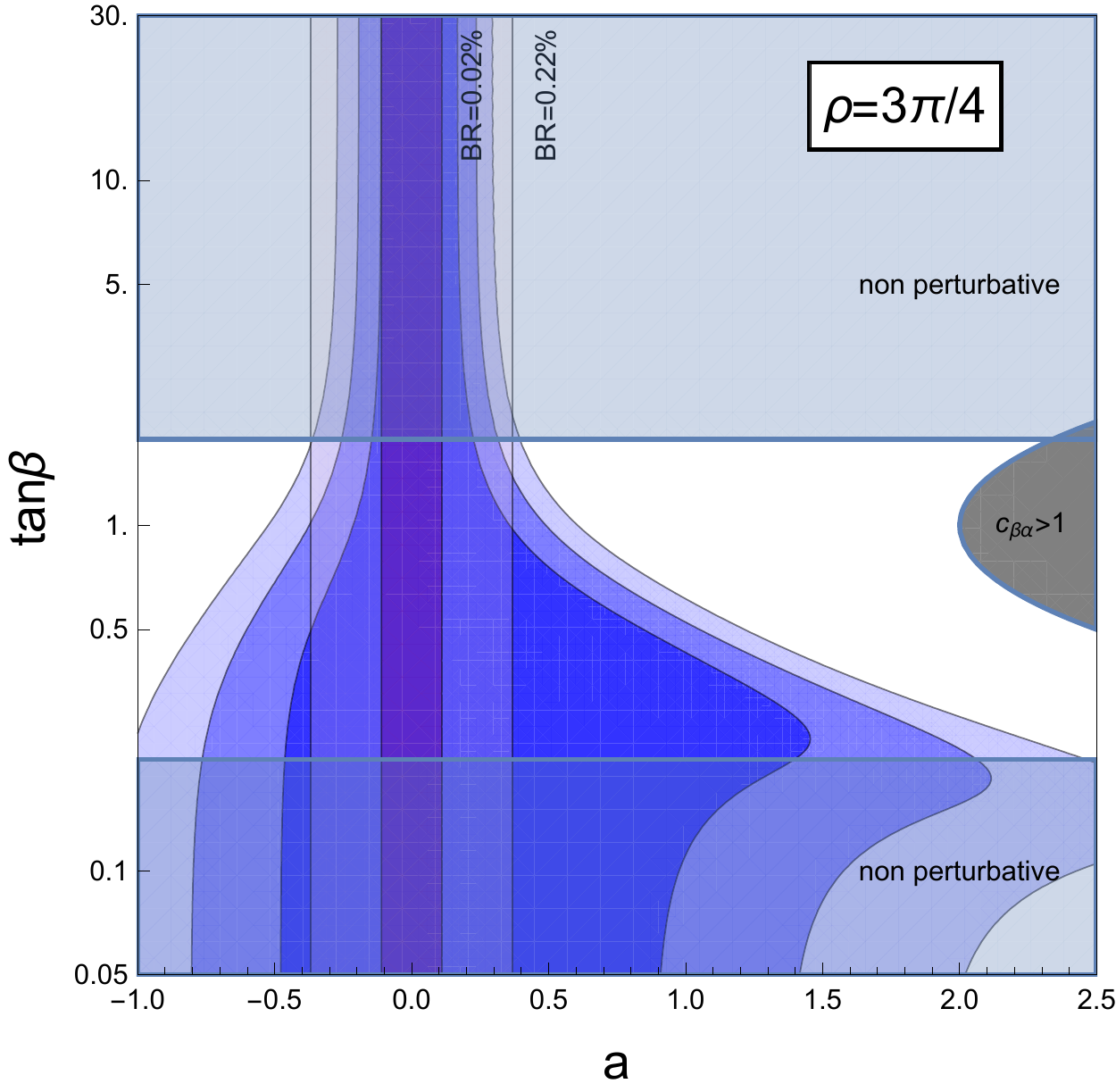}
\caption{Allowed parameter space in the $a$-$\tan\beta$ plane.  The left plot assumes $\rho=0$ (no mixing) while the right plot has $\rho=3\pi/4$, where top-charm flavor mixing effects are almost maximal.
Even though the mixing effect is maximized, it is difficult to realize a branching ratio of 0.22\% for the $t \to c h$ decay in the large $\tan\beta$ case. }
\label{fig:atanbeta}
\end{figure}

Fig.~\ref{fig:atanbeta} shows the allowed parameter space in the ($a$, $\tan \beta$) plane of our model, as constrained by the current Higgs data when the flavor mixing effect is switched off ($\rho=0$) in the left plot and switched on almost maximally ($\rho=3\pi/4$) in the right plot.  Again, the second branch of solution in the left plot appears only when $\tan\beta$ is sufficiently large.  At $95\%$ CL, the branch of $a\sim 0$ is constrained to have $|a| \alt 0.2$ when the mixing effect is switched off.
In the case with the mixing effect is larger, the allowed parameter region is slightly relaxed but 
still constrained to be $|a| \alt 0.3$ for $\tan \beta > 1$.  On the other hand, for the smaller $\tan \beta$ region, the allowed $a$ region is significantly extended.
Furthermore, one can see in the right plot that even with $\rho = 3\pi/4$, which provides almost maximized FCNC's in $\rho$, larger $\tan\beta$ is incompatible with $BR(t\to ch)=0.22\%$.  Therefore, we can conclude that there exists an upper bound on $\tan\beta$ for the nominal value of $BR(t\to ch)=0.22\%$~\footnote{With the exception of
the wrong Yukawa solution with a sufficiently large $\tan\beta$, although the parameters are compatible with the Higgs data at the $2\sigma$ level.}. 
Larger $\tan\beta$ with larger mixing $\rho$ is also disfavored by the perturbativity requirement.

\section{Further tests for the model}
\label{sec:confirmation}

Once we observe a sufficient number of $t \to ch$ events, it will be possible to check the chiral nature of the Higgs flavor-changing couplings as predicted in the model: the charm quark in the decay product should be right-handed.  
The spin analyzing power $\kappa_i$ of particle $i$ in the decay product is defined as 
\begin{eqnarray}
\frac{1}{\Gamma_i}\frac{d\Gamma_i}{d\cos\theta_i} = \frac{1}{2}( 1 + \kappa_i P \cos\theta_i) ~,
\end{eqnarray}
where $P$ is the polarization of the mother particle along a specific direction, called the polarization axis, $\Gamma_i$ is the partial decay width of the mode containing particle $i$, and 
$\theta_i$ is the polar angle of particle $i$ with respect to the polarization axis. The spin analyzing power of the charged lepton, $\kappa_{\ell^+}$, from the usual top decay $t \to b\ell^+\nu$ is known to have the largest value $+1$ at leading order~\cite{Bernreuther:2008ju}.  
We denote the spin analyzing power for the anti-top quark decay by $\bar{\kappa}$, and note that $\bar{\kappa}_{\bar{f}}= -\kappa_f$ assuming CP invariance.  Our model predicts $d\Gamma_{t \to ch}/d\cos\theta \propto 1+\cos\theta$, and the charm quark and the Higgs boson have the spin analyzing powers $\kappa_c=+1$ and $\kappa_h=-1$, respectively.  
Once we know the original top spin direction, we can readily determine $\kappa_{c}$ and $\kappa_h$ in the $t \to ch$ decay.

We have discussed in Ref.~\cite{Chiang:2015cba} the possibility of determining the chirality structure 
using the top spin correlation in $t\bar{t}$ production at the LHC~\cite{Aad:2014mfk, CMS:2015dva}
as the top quarks in top pair production are not polarized and not directly usable.
The differential cross section in the double theta distribution is given by:
\begin{eqnarray}
\frac{1}{\sigma} \frac{d\sigma}{d\cos\theta_i d\cos\theta_j} = \frac{1}{4}( 1 + A_{\rm hel} \, \kappa_i \bar{\kappa}_j \cos\theta_i \cos\theta_j) ~,
\end{eqnarray}
where $\theta_{i,j}$ are defined in the rest frame of the top and anti-top quarks, respectively, and 
the $t\bar{t}$ spin asymmetry defined in the helicity basis~\cite{Bernreuther:2004jv} is
\begin{eqnarray}
A_{\rm hel}
\equiv 
\frac{N(t_\uparrow \bar{t}_\uparrow) + N(t_\downarrow \bar{t}_\downarrow) - N(t_\uparrow \bar{t}_\downarrow) - N(t_\downarrow \bar{t}_\uparrow)}
{N(t_\uparrow \bar{t}_\uparrow) + N(t_\downarrow \bar{t}_\downarrow) + N(t_\uparrow \bar{t}_\downarrow) + N(t_\downarrow \bar{t}_\uparrow)}
\sim 0.35
\end{eqnarray}
at the LHC.

For a rough estimate of the required number of events to determine $\kappa_h$ (or $\kappa_c$) by measuring the angular distribution of $i=\ell^+$, $j=h$ and the corresponding anti-particle case, we have also introduced a simpler observable out of the above-mentioned observables as
\begin{eqnarray}
A_{\ell h} 
\equiv 
\frac{N( \cos\theta_\ell \cos\theta_h >0) - N( \cos\theta_\ell \cos\theta_h <0)}{N( \cos\theta_\ell \cos\theta_h >0) + N( \cos\theta_\ell \cos\theta_h <0)}=
\frac{A_{\rm hel} \kappa_{\ell^+} \bar{\kappa}_h}{4} \sim 0.088 \bar{\kappa}_h.
\end{eqnarray}

To confirm that $\kappa_h \sim -1$, we have to measure a positive $A_{\ell h}$ at a precision better than $0.088$.
The statistical uncertainty on $A_{\ell h}$ is then given by 
\begin{eqnarray}
\Delta A_{\ell h} \simeq \Delta N/N \simeq 1/\sqrt{N} < 0.088 ~,
\end{eqnarray}
implying that we need at least $\sim 130$ signal events to confirm the decay distribution structure at the $1\sigma$ level.
As we expect $3 \times 10^9$ top pair events using $\sigma(t\bar{t}) \sim 1$~nb and an integrated luminosity of 3000~fb$^{-1}$ at the 14-TeV LHC,
it provides $\sim 10^7$ $t\to ch$ events for the nominal branching ratio of 0.22\%.
Even considering only the cleanest mode $h\to \gamma\gamma$, we still expect $\sim 5000$ events after multiplying $BR(h\to \gamma\gamma) \sim 2.3 \times 10^{-3}$ and the leptonic decay branching ratio of the top quark. 
Besides, the $h \to b\bar{b}$ mode can be incorporated to enhance the signal significance~\cite{AguilarSaavedra:2004wm}. 
Therefore, with the assumption of the nominal branching ratio, one can easily determine the chirality structure of the flavor-changing Higgs coupling at the ultimate integrated luminosity in LHC Run-III.

Recently, the high energy upgrade of the LHC (HE-LHC) is more sceriously discussed and 
its target center of energy and integrated luminosity are realistically decided as 27~TeV and $12$~fb$^{-1}$~\cite{HELHCtalk}.
At $\sqrt{s}=27$~TeV, the $t\bar{t}$ cross section reaches 3.7 nb computed by Hathor 2.0~\cite{Aliev:2010zk}, 
therefore, we expect that the sensitivity on the branching ratio improves below 
$10^{-5}$ due to 15 times as many as $t\bar{t}$ events.

\section{conclusion}
\label{sec:conclusion}

In this Letter, we consider the current status of the top-specific variant axion model in light of the latest Higgs data and a slight excess in the $t\to ch$ decay .
This model is well-motivated to solve the strong CP and domain wall problems.  
As the top FCNC decay is one of the generic predictions of the model, we discuss whether it is possible to have a sizeable $BR(t \to ch)$ under various theoretical and phenomenological constraints.
We have found that to realize a rather large branching ratio of $0.22$\% in this model, $\tan\beta\sim 1$ or 
smaller is preferred by the current Higgs data.  Such a preference is also supported by the perturbativity requirement on the Yukawa couplings.  
In other words, what we have shown is that although the Higgs signal strength data are essentially the same as the SM predictions, our model can readily accommodate a sizable $BR(t\to ch)$ without conflicts with the Higgs data as long as $\tan \beta \lesssim 1$.

The $h$-$t$-$c$ vertex has a specific chirality structure according to the model. 
We therefore propose to measure this characteristic feature as an essential step toward verifying the model.
We have shown that this can be done by measuring the spin correlation in the top pair production through 
the $t \to ch$ mode and estimated that the required sensitivity for confirming the model can be achieved 
by the end of LHC Run-III, assuming $BR(t \to c h) = 0.22$\% is realized in Nature.

\section*{Acknowledgments}

This research was supported in part by the Ministry of Science and Technology of Taiwan under Grant No.\ NSC 100-2628-M-008-003-MY4 (C.-W.~C); and in part by the Grants-in-Aid for Scientific Research from the Ministry of Education, Culture, Sports, Science, and Technology (MEXT), Japan No.~26104009 and Grant-in-Aid No.~26287039 from the Japan Society for the Promotion of Science (JSPS) (T.~T.~Y.); 
JSPS Grant-in-Aid for Scientific Research Numbers JP16H03991, JP16H02176, 17H05399 (M.~T.);
and the World Premier International Research Center Initiative (WPI Initiative), MEXT, Japan (H.~F., M.~T. and T.~T.~Y.).

\end{document}